\documentclass[MSNbibl,nameyear,dvips]{arxstspdf}
\usepackage{tikz}
\usepackage{graphicx}
\usepackage{flushend}
\usepackage{stfloats}

\volume{29}
\issue{3}
\pubyear{2014}
\firstpage{363}
\lastpage{366}
\doi{10.1214/14-STS485} % kopijuoti is 'New paper accepted'
\referstodoi{10.1214/14-STS480}% pagrindinio straipsnio DOI, kai
\docsubty{FLA}

\makeatletter

\newcommand{\ind}{\mbox{$\,\perp\kern-5.5pt \perp\,$}}

\renewcommand{\citep}[1]{(\citeauthor{#1}, \citeyear{#1})}

\makeatother

\begin{document}
\begin{frontmatter}

\vspace*{12pt}\title{ACE Bounds; SEMs with Equilibrium Conditions}
\runtitle{ACE Bounds; SEMS with Equilibrium Conditions}%\thanksref{T1}
% kai straipsnis turi susijusiu diskusiju ir rejoinder'iu
%rejoinder at \relateddoi{r}{10.1214/00-STSXXXX}.}

\begin{aug}
\author[a]{\fnms{Thomas S.} \snm{Richardson}\ead[label=e1]{thomasr@uw.edu}}
\and
\author[b]{\fnms{James M.} \snm{Robins}\ead[label=e2]{robins@hsph.harvard.edu}}

\runauthor{T. S. Richardson and J. M. Robins}

\affiliation{University of Washington and Harvard School of Public Health}

\address[a]{Thomas S. Richardson is Professor and Chair, Department of Statistics, University of Washington,
Box 354322, Seattle, Washington 98195, USA \printead{e1}.}

\address[b]{James M. Robins is Mitchell L. and Robin LaFoley Dong Professor,
Department of Epidemiology, Harvard School of
Public Health,
677 Huntington Avenue, Boston, Massachusetts 02115, USA \printead{e2}.}
\end{aug}

% ABSTRACT

% KEYWORDS
\end{frontmatter}

We congratulate the author on an enlightening account of
the instrumental variable approach from the viewpoint of Econometrics.
We first make some comments regarding the bounds on the ACE under the
nonparametric IV model,
and then discuss potential outcomes in the market equilibrium model.

%s1 #&#
\section{ACE Bounds Under the IV Model}
\begin{figure}

\includegraphics{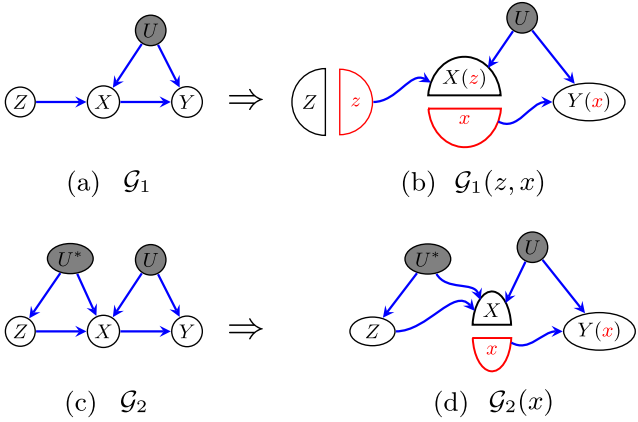}

\caption{\textup{(a)} IV model with no confounding between $Z$ and $X$; \textup{(b)}
SWIG representing $P(Z, X(z),Y(x),U)$;
\textup{(c)} IV model with confounding between $Z$ and $X$; \textup{(d)} SWIG
representing $P(Z, X,Y(x),U,U^*)$.}\label{figswig}
\end{figure}

We consider the model in which $X$ and $Y$ are binary, taking values in
$\{0,1\}$, while
$Z$ takes $K$ states $\{1,\ldots,K\}$. We use the notation $X(z_i)$ to indicate
$X(z = i)$, similarly $Y(x_j)$ for $Y(x = j)$. We consider four
different sets of assumptions:
\begin{longlist}[(iii)]
\item[(i)] $Z \ind Y({x}_0),Y({x}_1), X({z}_1),\ldots,X({z}_{K})$;
\item[(ii)] $Z \ind Y({x}_0),Y({x}_1)$;
\item[(iii)] for $i \in\{1,\ldots,K\}$, $j \in\{0,1\}$, $Z \ind
X(z_i),\linebreak[4] Y(x_j)$;
\item[(iv)] there exists a $U$ such that $U \ind Z$ and for $j \in\{
0,1\}$, $Y(x_j) \ind X,Z \mid U$.
\end{longlist}
Condition (i) is joint independence of $Z$ and all potential outcomes
for $Y$ and $X$. (ii) does not assume independence (or existence)
of counterfactuals for $X$.
(iii)~is a subset of the independences in (i), none of which involve
potential outcomes from different worlds.\footnote{In other words,
they do not involve both $Y(x_0)$ and $Y(x_1)$, nor $X(z_i)$ and
$X(z_j$) for $i\neq j$.}
The counterfactual independencies (i), (ii), (iii) arise most naturally
in the context
where the instrument is randomized, as depicted by the DAG in Figure~\ref{figswig}(a).
Assumption (iii) may be read (via d-separation) from the Single-World
Intervention Graph (SWIG)\footnote{See \citet{richardsonrobins2013}
for details.}
$\mathcal{G}_1(z,x)$, depicted in Figure~\ref{figswig}(b), which
represents the factorization of $P(Z,X(z),Y(x),U)$, implied by the IV model.

Lastly (iv) consists of only three independence statements, but does
assume the existence of an unobserved variable $U$ that
is sufficient to control for confounding between $X$ and $Y$. No
assumption is made concerning the existence of counterfactuals
$X(z)$; confounding variables ($U^*$) between $Z$ and $X$ are permitted
(so long as $U^* \ind U$). The DAG $\mathcal{G}_2$ and
corresponding SWIG $\mathcal{G}_2(x)$ are shown in Figure~\ref{figswig}(c), (d). In \citet{richardsonrobins2014}, we prove the following.

%th1 #&#
\begin{thm} Under any of the assumptions \textup{(i)}, \textup{(ii)}, \textup{(iii)}, \textup{(iv)}, the set of possible joint distributions
$P(Y(x_0), Y(x_1))$ are characterized by
the $8K$ inequalities:\vspace*{-2pt}
%
%e1 #&#
%e2 #&#
\begin{eqnarray}
\label{eqmarg}&&P\bigl(Y(x_i) = y\bigr) \nonumber\\[-1pt]
&&\quad \leq P(Y = y, X = i | Z = z)\\[-1pt]
&&\qquad {}+ P(X = 1-i |
Z = z),\nonumber
\\[4pt]
\label{eqjoint}&&P\bigl(Y(x_0) = y, Y(x_1) = \tilde{y}\bigr) \nonumber\\[-1pt]
&&\quad \leq P(Y
= y, X = 0 | Z = z)\\[-1pt]
&&\qquad {} + P(Y = \tilde{y}, X = 1 | Z = z).\nonumber
\end{eqnarray}
\end{thm}

Thus a distribution $P(X,Y | Z)$ is compatible with the stated
assumptions if and only if there exists a distribution
$P(Y(x_0), Y(x_1))$ satisfying (\ref{eqmarg}) and (\ref{eqjoint}).

%th2 #&#
\begin{thm} Under any of the assumptions \textup{(i)}, \textup{(ii)}, \textup{(iii)}, \textup{(iv)}
for all $i,j \in\{0,1\}$, $P(Y(x_i) = j) \leq
g(i,j)$, where\vspace*{-2pt}
{\fontsize{10.9}{12.9}\selectfont{\begin{eqnarray*}
g(i,j) &\equiv&\min \Bigl\{ \min_{z} \bigl[\vphantom{\hat{P}} P(X
= i, Y = j | Z = z)\\
&&\hphantom{\min \Bigl\{\min_{z} \bigl[}{} + P(X = 1-i | Z = z) \bigr],
\\
&&\hphantom{\min \Bigl\{} \min_{z, \tilde{z}: z \neq\tilde{z}} \bigl[ P(X = i, Y = j | Z
= z) \\
&&\hphantom{\min \Bigl\{\min_{z, \tilde{z}: z \neq\tilde{z}} \bigl[}{}+ P(X = 1-i, Y = 0 | Z = z)
\nonumber
\\
&&\hphantom{\min \Bigl\{\min_{z, \tilde{z}: z \neq\tilde{z}} \bigl[}{}  +  P(X = i, Y = j | Z =
\tilde{z}) \\
&&\hphantom{\min \Bigl\{\min_{z, \tilde{z}: z \neq\tilde{z}} \bigl[}{}+ P(X = 1-i, Y = 1 | Z = \tilde {z}) \bigr] \Bigr\}.
\nonumber
\end{eqnarray*}}}
Furthermore, $P(Y(x_0))$ and $P(Y(x_1))$ are variation independent.
Consequently,
\begin{eqnarray*}
1-g(1,0)-g(0,1) &\leq& \operatorname{ACE}(X \rightarrow Y)\\
& \leq& g(0,0)+g(1,1)-1.
\end{eqnarray*}
These bounds are sharp.
\end{thm}

%f1 #&#

Note that to evaluate $g(i,j)$ requires finding a minimum over $K^2$
expressions. In the case
where $K=2$, these bounds reduce to those given by \citet
{BalkPearboun1997}, who assume (i).\footnote{\citet{dawid2003} working
in a non-counterfactual framework also established the bounds for $K=2$
under the DAG in Figure~\ref{figswig}(a); however,
his proof also applies to Figure~\ref{figswig}(c). \citet
{robinsgreenland1996} observed that the Balke--Pearl bounds
were also sharp under (ii).} \citet{robins1989} and \citet{manski1990} derived what are called
the ``natural bounds'' on the ACE under the weaker assumption that $Z
\ind Y({x}_0)$ and $Z \ind Y({x}_1)$.
As noted by Imbens, without further assumptions these bounds are not
sharp. However, the natural bounds are sharp under (i) or (iii), if, in
addition, we assume there are no Defiers (an assumption that has
testable implications). \citet{chengsmall2006} considered bounds on
the ACE when $K=3$ under additional assumptions.
%

%s2 #&#
\section{Market Equilibrium and BiCausal Models}
Imbens' clear description of the market equilibrium model is
particularly informative.
We also strongly endorse the author's contention that the RHS of
systems of structural equations
should be interpreted as describing potential outcomes for the
LHS.\footnote{
\citet{pearl2000}, \citet{laucausal}, \citet{lauritzen02} argue that these are not
really ``equations''
but are better viewed as ``assignments'' in computer languages, for example, $ y
\leftarrow x +1$;
see also \citet{strotzwoldrecursive1960}, page 420.}

However, we note that this position has important implications both for
interpretation and inference.
Furthermore, it does not
seem to be universally accepted within Economics. \citet{leroy2006}
states that ``economic models use the equality symbol with its usual
mathematical meaning,
not with the meaning of the assignment operator'';
an approach that is clearly incompatible with an interpretation in
terms of potential outcomes. For example,
it becomes permissible to renormalize structural equations to change
which variable is on the LHS.

It has also been argued that statistical analyses of such models should be
invariant to the normalization; see \citet{hillier1990},
\citet{basmanncausal1963}.\hskip.2pt\footnote{For example, \citet{greene2003}, page 401, states (in the
context of the IV model):
``one significant virtue of [the Limited Information Maximum Likelihood
Estimator]
is its invariance to normalization of the equations.''}
Contrary to Imbens' remark,\footnote{Footnote 8, page 331.} this
alternative view does not appear to be motivated by
considerations of measurement error. \citet{leroy2006} makes clear
that he does not believe
that structural equations describe potential outcomes for endogenous
variables and does not discuss issues relating to measurement.\footnote
{For example,
\citet{leroy2006}, page 23, states that ``The assumption that it makes
sense to delete one or more of the structural equations
and replace the value of the internal variable so determined by a
constant without altering the other equations [\ldots] is virtually never satisfied
in economic models since each external variable typically affects
equilibrium values of more than one internal variable.''
He goes on to assert ``In fact, it is difficult to think of nontrivial
models in any area of research in which the [\ldots] assumption is satisfied.''}
Rather, this appears to be a fundamental difference in interpretation.

\begin{figure}

\includegraphics{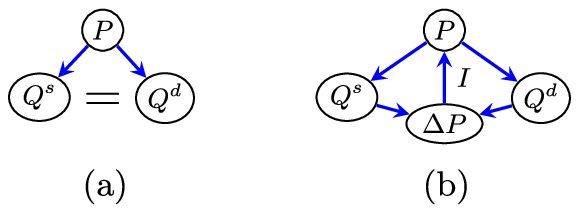}

\caption{\textup{(a)} Attempt to depict the bicausal model; \textup{(b)} a schematic
showing the deterministic system (\protect\ref{eqcon})--(\protect\ref{eqmer});
the edge
\protect\tikz\protect\path(0ex,0ex) edge[->] node[above=0pt, black]
{$\scriptscriptstyle I$} (3ex,0ex);
denotes that $P$ is the integral of $\Delta P$; see Iwasaki and Simon (\citeyear
{iwasaki1994}).}\label{figone}
\end{figure}

The market equilibrium model specifies potential outcomes for
$Q^d_t(p)$, $Q^s_t(p)$:
%
%e3 #&#
%e4 #&#
\begin{eqnarray}
\label{eqqd}Q^d_t(p)&=& \alpha^d + \beta^d p
+ \varepsilon_t^d,
\\
\label{eqqs}Q^s_t(p)&=& \alpha^s + \beta^s p
+ \varepsilon_t^s,
\end{eqnarray}
and imposes the equilibrium condition:\footnote{To simplify notation,
throughout we work directly in terms of $\log$ price and $\log$ quantity.}
%
%e5 #&#
\begin{eqnarray}
Q^d_t(p) = Q^s_t(p).\label{eqequ}
\end{eqnarray}
\citet{strotzwoldrecursive1960} described such systems as \textit
{bicausal}.
It should be observed that the model does not
specify potential outcomes for price ($P_t(q_s,q_d)$), nor does it view
price as externally determined (i.e., exogenous).
Instead price is determined implicitly as a consequence of the
equilibrium condition. In this regard, the
model might be regarded as incomplete: Indeed \citet
{haavelmowhat1958} is quite critical of this model for failing to
offer any \emph{explanation} as to
how the equilibrium price is determined. The model also falls outside
the scope of non-parametric structural equation models (NPSEM) (see, e.g., \cite{pearl2000}), which require one equation for each
endogenous variable;\footnote{%
Indeed \citet{leroy2006} argues against the interpretation of
structural equations in terms of potential outcomes on the grounds that
this interpretation, as advanced by Pearl, requires a one-to-one
mapping between equations and endogenous variables that he argues, does not
make sense for the market equilibrium model.}
likewise the model
defies standard graphical representation, though see Figure~\ref{figone}(a).
%
%f2 #&#

A related question concerns whether there exist dynamic acyclic (i.e.,
recursive)
systems of structural equations
that lead to the equilibrium distribution corresponding either to a
cyclic system
of structural equations or a bicausal system.\footnote{%
Analysis of this question was stimulated by a heated debate that arose between
Wold, who advocated a recursive, regression-based approach to demand
analysis, and Haavelmo
and the Cowles Commission who advocated simultaneous equations. See
\citet{haavelmostatistical1943}, \citet{woldbentzelstatistical1946},
\citet{woldjureendemand1953}, \citet{bentzelhansenrecursiveness1954},
\citet{strotzwoldrecursive1960}, \citet{basmanncausal1963}; historical overviews
are given by \citet{morganstamping1991},
\citet{epsteinhistory1987}.} \citet{fishcorr} provides just such a ``correspondence principle''
under which the distribution implied by a cyclic linear SEM is obtained
as a time average of a deterministic
set of first order difference equations reaching a static equilibrium
subject to stochastic boundary conditions.
The correspondence assumes that the equilibration time is very fast
relative to the interval between observations
so the time averaged variables are in deterministic equilibrium.
Fisher also derived conditions on the coefficient matrices of a cyclic
SEM that are required in order for the system
to reach equilibrium; in fact he further required that each subset of
structural equations also have this property.

However, Fisher's correspondence presumes a normalization under which
each variable is associated with
a single equation (as in an NPSEM), and hence would not apply to a
bicausal system. \citet{richphd}, Chapter~2,
described a system of finite difference equations that gives rise to
the bicausal system~(\ref{eqqd})--(\ref{eqequ}):
%
%e6 #&#
%e7 #&#
%e8 #&#
\begin{eqnarray}
\label{eqcon}\mbox{Consumers:}&&\hspace*{4pt}   Q^d_{t+(k+1)\delta}(p_{t+k\delta})\nonumber \\[-8pt]\\[-8pt]
&&\hspace*{4pt}\quad =
\alpha^d+ \beta^d p_{t+k\delta} + \varepsilon_{t}^d,\nonumber
\\
\label{eqsup}\mbox{Suppliers:}&&\hspace*{4pt}  Q^s_{t+(k+1)\delta}(p_{t+k\delta}) \nonumber\\[-8pt]\\[-8pt]
&&\hspace*{4pt}\quad =
\alpha^s+ \beta^s p_{t+k\delta} + \varepsilon_{t}^s,\nonumber
\\
\label{eqmer}\mbox{Merchants:}&&\hspace*{4pt}  P_{t+(k+1)\delta}\bigl(q^d_{t+k\delta},
q^s_{t+k\delta},p_{t+k\delta}\bigr) \nonumber\\[-8pt]\\[-8pt]
&&\hspace*{4pt}\quad = p_{t+k\delta} +
\lambda \bigl(q^d_{t+k\delta} - q^s_{t+k\delta}
\bigr),\nonumber
\end{eqnarray}
for $k=\{0,\ldots, \delta^{-1}-1\}$. Note that the disturbances
$(\varepsilon_{t}^d, \varepsilon_{t}^s)$ represent boundary
conditions and hence
remain fixed during the interval $[t,t+1)$.
As in Fisher's correspondence, the observed variables correspond to
limiting time-averages over a unit
interval:
\begin{eqnarray*}
\overline{Q}^d_t &=& \lim_{\delta\rightarrow0} \delta
\sum_{k=0}^{\delta^{-1}-1} {Q}^d_{t+k\delta},\quad
\overline{Q}^s_t = \lim_{\delta\rightarrow0} \delta
\sum_{k=0}^{\delta^{-1}-1} {Q}^s_{t+k\delta},\\
 \overline{P}_t &=& \lim_{\delta\rightarrow0} \delta\sum
_{k=0}^{\delta^{-1}-1} {P}_{t+k\delta}.
\end{eqnarray*}
Under suitable conditions on the coefficients, $(\overline{Q}^d_t,
\overline{Q}^s_t,\allowbreak  \overline{P}_t)$ obey
equations (\ref{eqqd})--(\ref{eqequ}). Note that Merchants'
equation (\ref{eqmer}) which includes $P$, leads to the equilibrium
condition (\ref{eqequ}) that does not.\footnote{In causal terms,
this model is similar to one presented in \citet{wold1959}. Wold
viewed his model as a formalization of Cournot's theories.}
It might be objected to the proposed model that there is no disturbance
term in equation (\ref{eqmer}).
The explanation for this is that the disturbance terms in the
nonrecursive model correspond to constant factors in the deterministic
evolution.
The equation for price gives the change in price during a small
interval (length $\delta$) to the discrepancy between supply and demand.
Adding a disturbance term would say that throughout the observation
period (length $1$) the Merchants' reaction to change in price was off
by a constant factor, so that even if quantities supplied and demanded
were identical, the Merchants would change the price. Thus, if we add
an error $\varepsilon^p_t$ the model will not, in general, arrive at
equilibrium
within the unit interval.\footnote{Having said this, the equations
(\ref{eqcon}) and (\ref{eqsup}) still imply that producers and
consumers make systematic errors in computing prices over a time-scale
of length $\delta$.}

\citet{iwasaki1994} represent equilibrating mechanisms via ``causal
influence diagrams'' in which the derivatives of variables are included.
Under this scheme, model (\ref{eqcon})--(\ref{eqmer}) is
represented by the graph in Figure~\ref{figone}(b).
This example serves to show that time averages of (deterministic)
equilibrating systems need not have a structural equation for each variable.
See also \citep{2001.dash.esqaru} for related work.

\section*{Acknowledgments}
This work was supported by the US National Institutes of Health Grant
R01 AI032475; Richardson was also supported by the US National Science
Foundation Grant CNS-0855230.

%suskaldyti doi

% imsref loaded by jurgita.kaciuliene, 2014-07-23 13:00:59


\begin{thebibliography}{28}
% pybtex-1.13. Style name=ims, version=2.8, label_style=nameyear, sorting_style=complex, cfg=None, language=None.


%b1 ###
%b1 #&#
\bibitem[\protect\citeauthoryear{Balke and Pearl}{1997}]{BalkPearboun1997}
\begin{barticle}[author]
\bauthor{\bsnm{Balke},~\bfnm{Alexander}\binits{A.}} \AND
\bauthor{\bsnm{Pearl},~\bfnm{Judea}\binits{J.}}
(\byear{1997}).
\btitle{Bounds on treatment effects from studies with imperfect compliance}.
\bjournal{J. Amer. Statist. Assoc.}
\bvolume{92}
\bpages{1171--1176}.
\end{barticle}
\bptok{imsref}%
\endbibitem

%b2 ###
%b2 #&#
\bibitem[\protect\citeauthoryear{Basmann}{1963}]{basmanncausal1963}
\begin{barticle}[author]
\bauthor{\bsnm{Basmann},~\bfnm{R.~L.}\binits{R.~L.}}
(\byear{1963}).
\btitle{The causal interpretation of non-triangular systems of economic relations (with discussion)}.
\bjournal{Econometrica}
\bvolume{31}
\bpages{439--453}.
\end{barticle}
\bptok{imsref}%
\endbibitem

%b3 ###
%b3 #&#
\bibitem[\protect\citeauthoryear{Bentzel and Hansen}{1954}]{bentzelhansenrecursiveness1954}
\begin{barticle}[author]
\bauthor{\bsnm{Bentzel},~\bfnm{R.}\binits{R.}} \AND
\bauthor{\bsnm{Hansen},~\bfnm{B.}\binits{B.}}
(\byear{1954}).
\btitle{On recursiveness and interdependency in economic models}.
\bjournal{Rev. Econom. Stud.}
\bvolume{22}
\bpages{153--168}.
\end{barticle}
\bptok{imsref}%
\endbibitem

%b4 ###
%b4 #&#
\bibitem[\protect\citeauthoryear{Bentzel and Wold}{1946}]{woldbentzelstatistical1946}
\begin{barticle}[mr]
\bauthor{\bsnm{Bentzel},~\bfnm{R.}\binits{R.}} \AND
\bauthor{\bsnm{Wold},~\bfnm{H.}\binits{H.}}
(\byear{1946}).
\btitle{On statistical demand analysis from the viewpoint of simultaneous equations}.
\bjournal{Skand. Aktuarietidskr.}
\bvolume{29}
\bpages{95--114}.
\bid{mr={0017907}}
\end{barticle}
\bptok{imsref}%
\endbibitem

%b5 ###
%b5 #&#
\bibitem[\protect\citeauthoryear{Cheng and Small}{2006}]{chengsmall2006}
\begin{barticle}[mr]
\bauthor{\bsnm{Cheng},~\bfnm{Jing}\binits{J.}} \AND
\bauthor{\bsnm{Small},~\bfnm{Dylan~S.}\binits{D.~S.}}
(\byear{2006}).
\btitle{Bounds on causal effects in three-arm trials with non-compliance}.
\bjournal{J. R. Stat. Soc. Ser. B Stat. Methodol.}
\bvolume{68}
\bpages{815--836}.
\bid{doi={10.1111/j.1467-9868.2006.00568.x}, issn={1369-7412}, mr={2301296}}
\end{barticle}
\bptok{imsref}%
% NOT OUTPUTED:
%   issn = 1369-7412
%   url = http://dx.doi.org/10.1111/j.1467-9868.2006.00568.x
%   number = 5
%   fjournal = Journal of the Royal Statistical Society. Series B. Statistical Methodology
\endbibitem


\bibitem[\protect\citeauthoryear{Dash and Druzdzel}{2001}]{2001.dash.esqaru}
%
\begin{binproceedings}[author]
\bauthor{\bsnm{Dash}, \bfnm{Denver}\binits{D.}} \AND
\bauthor{\bsnm{Druzdzel}, \bfnm{Marek J.}\binits{M. J.}}
(\byear{2001}).
\btitle{Caveats for causal reasoning with equilibrium models}.
In \bbooktitle{Proceedings of the Sixth European Conference on
Symbolic and Quantitative Approaches\vadjust{\goodbreak} to Reasoning with Uncertainty (ECSQARU), Toulouse, France}
(\beditor{\bfnm{Salem}\binits{S.} \bsnm{Benferhat}} \AND
\beditor{\bfnm{Philippe}\binits{P.} \bsnm{Besnard}}, eds.).
\bseries{Lecture Notes in Artificial Intelligence}
\bvolume{2143}
\bpages{192--203}.
\bpublisher{Springer},
\blocation{Berlin}.
\end{binproceedings}
\bptok{imsref}%
%
% NOT OUTPUTED:
% url = http://ceur-ws.org/Vol-818/paper3.pdf
\endbibitem





%b6 ###
%b6 #&#
\bibitem[\protect\citeauthoryear{Dawid}{2003}]{dawid2003}
\begin{bincollection}[mr]
\bauthor{\bsnm{Dawid},~\bfnm{A.~Philip}\binits{A.~P.}}
(\byear{2003}).
\btitle{Causal inference using influence diagrams: The problem of partial compliance}.
In \bbooktitle{Highly Structured Stochastic Systems}
(\beditor{\bfnm{P. J.}\binits{P. J.} \bsnm{Green}},
\beditor{\bfnm{N. L.}\binits{N. L.} \bsnm{Hjort}} \AND
\beditor{\bfnm{S.}\binits{S.} \bsnm{Richardson}}, eds.).
\bseries{Oxford Statist. Sci. Ser.}
\bvolume{27}
\bpages{45--81}.
\bpublisher{Oxford Univ. Press},
\blocation{Oxford}.
\bid{mr={2082406}}
\bptnote{check related}%
\end{bincollection}
\bptok{imsref}%
\endbibitem

%b7 ###
%b7 #&#
\bibitem[\protect\citeauthoryear{Epstein}{1987}]{epsteinhistory1987}
\begin{bbook}[mr]
\bauthor{\bsnm{Epstein},~\bfnm{Roy~J.}\binits{R.~J.}}
(\byear{1987}).
\btitle{A History of Econometrics}.
\bseries{Contributions to Economic Analysis}
\bvolume{165}.
\bpublisher{North-Holland},
\blocation{Amsterdam}.
\bid{mr={0918969}}
\end{bbook}
\bptok{imsref}%
% NOT OUTPUTED:
%   isbn = 0-444-70267-9
%   fpage = x+254
\endbibitem

%b8 ###
%b8 #&#
\bibitem[\protect\citeauthoryear{Fisher}{1970}]{fishcorr}
\begin{barticle}[author]
\bauthor{\bsnm{Fisher},~\bfnm{F.~M.}\binits{F.~M.}}
(\byear{1970}).
\btitle{A correspondence principle for simultaneous equation models}.
\bjournal{Econometrica}
\bvolume{38}
\bpages{73--92}.
\end{barticle}
\bptok{imsref}%
\endbibitem

%b9 ###
%b9 #&#
\bibitem[\protect\citeauthoryear{Greene}{2003}]{greene2003}
\begin{bbook}[author]
\bauthor{\bsnm{Greene},~\bfnm{William~H.}\binits{W.~H.}}
(\byear{2003}).
\btitle{Econometric Analysis},
\bedition{5th} ed.
\bpublisher{Prentice Hall},
\blocation{Upper Saddle River, NJ}.
\end{bbook}
\bptok{imsref}%
\endbibitem

%b10 ###
%b10 #&#
\bibitem[\protect\citeauthoryear{Haavelmo}{1943}]{haavelmostatistical1943}
\begin{barticle}[mr]
\bauthor{\bsnm{Haavelmo},~\bfnm{Trygve}\binits{T.}}
(\byear{1943}).
\btitle{The statistical implications of a system of simultaneous equations}.
\bjournal{Econometrica}
\bvolume{11}
\bpages{1--12}.
\bid{issn={0012-9682}, mr={0007954}}
\end{barticle}
\bptok{imsref}%
% NOT OUTPUTED:
%   issn = 0012-9682
%   fjournal = Econometrica. Journal of the Econometric Society
\endbibitem

%b11 ###
%b11 #&#
\bibitem[\protect\citeauthoryear{Haavelmo}{1958}]{haavelmowhat1958}
\begin{barticle}[author]
\bauthor{\bsnm{Haavelmo},~\bfnm{T.}\binits{T.}}
(\byear{1958}).
\btitle{Hva kan statiske likevektsmodeller fortelle oss?}
\bjournal{National{\o}konomisk Tidsskrift}
\bvolume{96 (Suppl.)}
\bpages{138--145}
\bnote{(in {N}orwegian). English translation published as: {W}hat can static equilibrium models tell us? \textit{Economic Inquiry} \textbf{12} (1974) 27--34}.
\end{barticle}
\bptok{imsref}%
\endbibitem

%b12 ###
%b12 #&#
\bibitem[\protect\citeauthoryear{Hillier}{1990}]{hillier1990}
\begin{barticle}[mr]
\bauthor{\bsnm{Hillier},~\bfnm{Grant~H.}\binits{G.~H.}}
(\byear{1990}).
\btitle{On the normalization of structural equations: Properties of direction estimators}.
\bjournal{Econometrica}
\bvolume{58}
\bpages{1181--1194}.
\bid{doi={10.2307/2938305}, issn={0012-9682}, mr={1079413}}
\end{barticle}
\bptok{imsref}%
% NOT OUTPUTED:
%   issn = 0012-9682
%   url = http://dx.doi.org/10.2307/2938305
%   number = 5
%   coden = ECMTA7
%   fjournal = Econometrica. Journal of the Econometric Society
\endbibitem

%b13 ###
%b13 #&#
\bibitem[\protect\citeauthoryear{Iwasaki and Simon}{1994}]{iwasaki1994}
\begin{barticle}[mr]
\bauthor{\bsnm{Iwasaki},~\bfnm{Yumi}\binits{Y.}} \AND
\bauthor{\bsnm{Simon},~\bfnm{Herbert~A.}\binits{H.~A.}}
(\byear{1994}).
\btitle{Causality and model abstraction}.
\bjournal{Artificial Intelligence}
\bvolume{67}
\bpages{143--194}.
\bid{doi={10.1016/0004-3702(94)90014-0}, issn={0004-3702}, mr={1281657}}
\end{barticle}
\bptok{imsref}%
% NOT OUTPUTED:
%   issn = 0004-3702
%   url = http://dx.doi.org/10.1016/0004-3702(94)90014-0
%   number = 1
%   coden = AINTBB
%   fjournal = Artificial Intelligence
\endbibitem

%b14 ###
%b14 #&#
\bibitem[\protect\citeauthoryear{Lauritzen}{2001}]{laucausal}
\begin{bincollection}[mr]
\bauthor{\bsnm{Lauritzen},~\bfnm{Steffen~L.}\binits{S.~L.}}
(\byear{2001}).
\btitle{Causal inference from graphical models}.
In \bbooktitle{Complex Stochastic Systems ({E}indhoven, 1999)}.
\bseries{Monogr. Statist. Appl. Probab.}
\bvolume{87}
\bpages{63--107}.
\bpublisher{Chapman \& Hall/CRC},
\blocation{Boca Raton, FL}.
\bid{mr={1893411}}
\end{bincollection}
\bptok{imsref}%
\endbibitem

%b15 ###
%b15 #&#
\bibitem[\protect\citeauthoryear{Lauritzen and Richardson}{2002}]{lauritzen02}
\begin{barticle}[mr]
\bauthor{\bsnm{Lauritzen},~\bfnm{Steffen~L.}\binits{S.~L.}} \AND
\bauthor{\bsnm{Richardson},~\bfnm{Thomas~S.}\binits{T.~S.}}
(\byear{2002}).
\btitle{Chain graph models and their causal interpretations}.
\bjournal{J. R. Stat. Soc. Ser. B Stat. Methodol.}
\bvolume{64}
\bpages{321--361}.
\bid{doi={10.1111/1467-9868.00340}, issn={1369-7412}, mr={1924296}}
\bptnote{check related}%
\end{barticle}
\bptok{imsref}%
% NOT OUTPUTED:
%   issn = 1369-7412
%   url = http://dx.doi.org/10.1111/1467-9868.00340
%   number = 3
%   fjournal = Journal of the Royal Statistical Society. Series B. Statistical Methodology
\endbibitem

%b16 ###
%b16 #&#
\bibitem[\protect\citeauthoryear{{LeRoy}}{2006}]{leroy2006}
\begin{btechreport}[author]
\bauthor{\bsnm{{LeRoy}},~\bfnm{Stephen~F.}\binits{S.~F.}}
(\byear{2006}).
\btitle{Causality in economics}.
\btype{Technical report, Univ. California, Santa Barbara}.
\end{btechreport}
\bptok{imsref}%
% NOT OUTPUTED:
%   publisher = UC Santa Barbara
\endbibitem

%b17 ###
%b17 #&#
\bibitem[\protect\citeauthoryear{Manski}{1990}]{manski1990}
\begin{barticle}[author]
\bauthor{\bsnm{Manski},~\bfnm{C.~F.}\binits{C.~F.}}
(\byear{1990}).
\btitle{Non-parametric bounds on treatment effects}.
\bjournal{American Economic Review}
\bvolume{80}
\bpages{351--374}.
\end{barticle}
\bptok{imsref}%
\endbibitem

%b18 ###
%b18 #&#
\bibitem[\protect\citeauthoryear{Morgan}{1991}]{morganstamping1991}
\begin{bincollection}[author]
\bauthor{\bsnm{Morgan},~\bfnm{M.~S.}\binits{M.~S.}}
(\byear{1991}).
\btitle{The stamping out of process analysis in econometrics}.
In \bbooktitle{Appraising Economic Theories}
(\beditor{\bfnm{Neil}\binits{N.}~\bparticle{de}~\bsnm{Marchi}} \AND
\beditor{\bfnm{Mark}\binits{M.}~\bsnm{Blaug}}, eds.)
\bpages{237--272}.
\bpublisher{Edward Elgar},
\blocation{Cheltenham}.
\end{bincollection}
\bptok{imsref}%
\endbibitem

%b19 ###
%b19 #&#
\bibitem[\protect\citeauthoryear{Pearl}{2000}]{pearl2000}
\begin{bbook}[mr]
\bauthor{\bsnm{Pearl},~\bfnm{Judea}\binits{J.}}
(\byear{2000}).
\btitle{Causality: Models, Reasoning, and Inference}.
\bpublisher{Cambridge Univ. Press},
\blocation{Cambridge}.
\bid{mr={1744773}}
\end{bbook}
\bptok{imsref}%
% NOT OUTPUTED:
%   isbn = 0-521-77362-8
%   fpage = xvi+384
\endbibitem

%b20 ###
%b20 #&#
\bibitem[\protect\citeauthoryear{Richardson}{1996}]{richphd}
\begin{bphdthesis}[author]
\bauthor{\bsnm{Richardson},~\bfnm{T.~S.}\binits{T.~S.}}
(\byear{1996}).
\btitle{Models of feedback: Interpretation and discovery}.
\btype{Ph.D. thesis, Carnegie-Mellon Univ}.
\end{bphdthesis}
\bptok{imsref}%
% NOT OUTPUTED:
%   publisher =
\endbibitem

%b21 ###
%b21 #&#
\bibitem[\protect\citeauthoryear{Richardson and Robins}{2013}]{richardsonrobins2013}
\begin{btechreport}[author]
\bauthor{\bsnm{Richardson},~\bfnm{Thomas~S.}\binits{T.~S.}} \AND
\bauthor{\bsnm{Robins},~\bfnm{James~M.}\binits{J.~M.}}
(\byear{2013}).
\btitle{{S}ingle {W}orld {I}ntervention {G}raphs {(SWIGs)}: A unification of the counterfactual and graphical approaches to
causality}.
\btype{Technical Report 128, Center for Statistics and the Social Sciences, Univ.
Washington, Seattle, WA}.
\end{btechreport}
\bptok{imsref}%
% NOT OUTPUTED:
%   publisher = Center for Statistics and the Social Sciences, Univ. Washington
\endbibitem

%b22 ###
%b22 #&#
\bibitem[\protect\citeauthoryear{Richardson and Robins}{2014}]{richardsonrobins2014}
\begin{bunpublished}[author]
\bauthor{\bsnm{Richardson},~\bfnm{T.~S.}\binits{T.~S.}} \AND
\bauthor{\bsnm{Robins},~\bfnm{J.~M.}\binits{J.~M.}}
(\byear{2014}).
\btitle{Assumptions and bounds in the instrumental variable model}.
\bnote{Preprint}.
\end{bunpublished}
\bptok{imsref}%
\endbibitem

%b23 ###
%b23 #&#
\bibitem[\protect\citeauthoryear{Robins}{1989}]{robins1989}
\begin{bincollection}[author]
\bauthor{\bsnm{Robins},~\bfnm{J.~M.}\binits{J.~M.}}
(\byear{1989}).
\btitle{The analysis of randomized and non-randomized AIDS treatment trials using a new approach to causal inference in longitudinal studies.}
In \bbooktitle{Health Service Research Methodology: A Focus on {AIDS}}
(\beditor{\bfnm{L.}\binits{L.}~\bsnm{Sechrest}},
\beditor{\bfnm{H.}\binits{H.}~\bsnm{Freeman}} \AND
\beditor{\bfnm{A.}\binits{A.}~\bsnm{Mulley}}, eds.).
\bpublisher{U.S. Public Health Service},
\blocation{Washington, DC}.
\end{bincollection}
\bptok{imsref}%
\endbibitem

%b24 ###
%b24 #&#
\bibitem[\protect\citeauthoryear{Robins and Greenland}{1996}]{robinsgreenland1996}
\begin{barticle}[author]
\bauthor{\bsnm{Robins},~\bfnm{James~M.}\binits{J.~M.}} \AND
\bauthor{\bsnm{Greenland},~\bfnm{Sander}\binits{S.}}
(\byear{1996}).
\btitle{Identification of causal effects using instrumental variables: Comment}.
\bjournal{J. Amer. Statist. Assoc.}
\bvolume{91}
\bpages{456--458}.
\bid{issn={01621459}}
\end{barticle}
\bptok{imsref}%
\endbibitem

%b25 ###
%b25 #&#
\bibitem[\protect\citeauthoryear{Strotz and Wold}{1960}]{strotzwoldrecursive1960}
\begin{barticle}[mr]
\bauthor{\bsnm{Strotz},~\bfnm{Robert~H.}\binits{R.~H.}} \AND
\bauthor{\bsnm{Wold},~\bfnm{H.~O.~A.}\binits{H.~O.~A.}}
(\byear{1960}).
\btitle{Recursive vs. nonrecursive systems: An attempt at synthesis}.
\bjournal{Econometrica}
\bvolume{28}
\bpages{417--427}.
\bid{issn={0012-9682}, mr={0120034}}
\end{barticle}
\bptok{imsref}%
% NOT OUTPUTED:
%   issn = 0012-9682
%   fjournal = Econometrica. Journal of the Econometric Society
\endbibitem\vfill\eject

%b26 ###
%b26 #&#
\bibitem[\protect\citeauthoryear{Wold}{1959}]{wold1959}
\begin{bincollection}[mr]
\bauthor{\bsnm{Wold},~\bfnm{Herman~O.~A.}\binits{H.~O.~A.}}
(\byear{1959}).
\btitle{Ends and means in econometric model building}.
In \bbooktitle{Probability and Statistics: {T}he {H}arald {C}ram\'er Volume} ({U}. {G}renander, ed.)
\bpages{355--434}.
\bpublisher{Almqvist \& Wiksell},
\blocation{Stockholm}.
\bid{mr={0109088}}
\end{bincollection}
\bptok{imsref}%
\endbibitem

%b27 ###
%b27 #&#
\bibitem[\protect\citeauthoryear{Wold and Jur{\'e}en}{1953}]{woldjureendemand1953}
\begin{bbook}[author]
\bauthor{\bsnm{Wold},~\bfnm{H.~O.~A.}\binits{H.~O.~A.}} \AND
\bauthor{\bsnm{Jur{\'e}en},~\bfnm{L.}\binits{L.}}
(\byear{1953}).
\btitle{Demand Analysis}.
\bpublisher{Wiley},
\blocation{New York}.
\end{bbook}
\bptok{imsref}%
\endbibitem\vfill


\end{thebibliography}
\end{document}